# A STUDY ON THE USAGE OF DATA STRUCTURES IN INFORMATION RETRIEVAL

V. R. Kanagavalli[1], G. Maheeja[2]

1- Associate professor, Department of Computer Applications, Sri Sai Ram Engineering College
2- PG student, Department of Computer Applications, Sri Sai Ram Engineering College

**Abstract:**

This paper tries to throw light in the usage of data structures in the field of information retrieval. Information retrieval is an area of study which is gaining momentum as the need and urge for sharing and exploring information is growing day by day. Data structures have been the area of research for a long period in the arena of computer science. The need to have efficient data structures has become even more important as the data grows in an exponential nature.

## 1. Introduction

Data has always been and is becoming a resource that needs to be judiciously used and shared for the benefit of the organizations and institutions. More than ever, there is a massive data sharing these days with the increasing technological updates and social networking sites. The processed data is called as information and the task of finding them from the existing repository are called as information retrieval. Information retrieval is an area of research actively involving multiple domains of computer science. Information retrieval exploits many of the fundamental concepts of computer science and is used as a tool in various advanced research areas of computer science. For an example, the field of text mining uses information retrieval as a first step for before applying other mining operations.

## 2. Information retrieval

### 2.1 Information retrieval and information extraction

Information retrieval (IR) and information extraction (IE) are the terms that are often used interchanged by mistake. They are thoroughly different areas with unique and different tasks as the goal. Information extraction does not have any goal or target to be achieved. It does use templates for providing a structure to an otherwise unstructured information. Information retrieval requires sophisticated techniques as it needs to satisfy the user requirement of finding the exact information from an existing repository. Information retrieval also involves auxiliary functions of selecting an apt index for better querying and an intelligent information retrieval system uses the feedback from the user to build upon the existing system and to fine tune the techniques. Information retrieval has functions identical to data mining such as summarization and clustering.

The Figure 1 depicts the general architecture of an information retrieval system.

### 2.2 Performance of an IR system

The success rate or performance of an information retrieval system is calculated based on the response time of the system and also the quality of the output. The quality of the response from information retrieval is again a qualitative term which can be judged by the feedback from the user. Usually the metrics used for the quality measurement are precision and recall. The recall measure is defined as the percentage of relevant documents retrieved to the total number of relevant documents. The precision measure is the percentage of relevant documents retrieved to the total number of retrieved documents. Though these metrics have precise definitions the relevance of the documents is again a qualitative term.

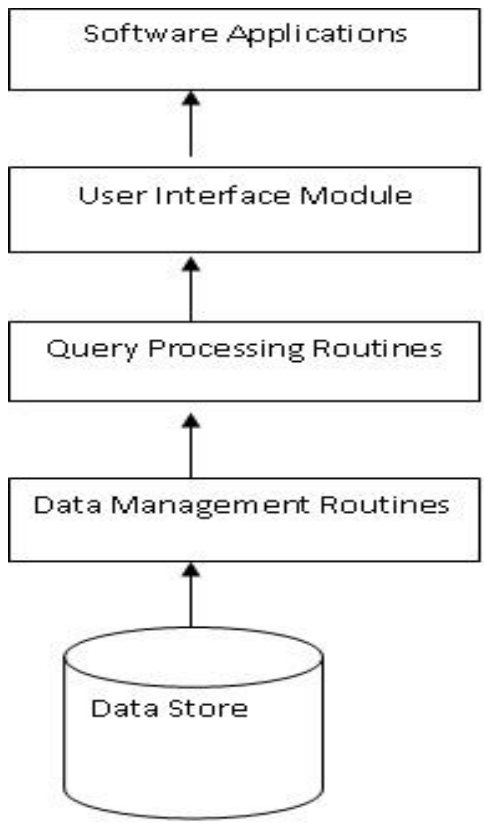

**Figure 1. Architecture of a typical IR system**

The response time of the information retrieval system is a quantitative metric as it is measurable. The factors that affects the response time of the information retrieval system are the size and arrangement of the corpus to be searched, the type of the index used, and the type of the query posed to the system. So when we need to decrease the response time of the IR system we need to concentrate on the type and size of corpus, type of index, and the type of the query along with the searching technique used. Now we consider the deployment of data structures in the field of information retrieval and how the choice of the data structures affects the performance of the information retrieval system.

3. **Data structures**

Data structures are the different techniques used to store the data in the persistent memory. The purpose of the data structures vary in each application. In [3], the authors classify the data structures based on the purpose for which it is used. The data structures like arrays, linked structures, hash tables are primarily used for storing the data and hence are classified as storage structures. The other category of data structures such as stacks, queues and priority queues are used for processing the data and these are classified as process-oriented data structures. There is still some data structures left out, which not only simply store the data but yield to the description of the data by the way it is arranged in it. Collections, sets, linear lists, binary trees, etc. are the data structures which describe the nature of the data held in it and hence the authors call it descriptive data structures.

4. **Information retrieval tasks**

The primary purpose of information retrieval is to answer the query posed by the users. Researchers are working towards not only satisfying the queries posed by the users but also to predict the queries that would be posed by the users to a document corpus. Fei Song and Bruce Croft in [2] compute the probabilities of generating the user's query terms with respect to each of the documents present in the corpus. The effectiveness of any information retrieval system is based on the feedback from the user. In [4] the authors evaluate the performance of the information retrieval based on the satisfaction of the user preferences.

5. **Storage Data structures in Information retrieval**

Information retrieval uses word oriented indexing techniques for retrieving the documents based on the query from the users. There are different types of indexing structures used namely signature files [7], inverted files etc., but primarily they employ the hash function and hash tables.

A hash data structure links key values to the data items. A hash function is used to map the search key to a key value. The key value usually denotes the bucket number to which the data item belongs. A bucket is nothing but a memory area. A hash table is most effective than most of the array structures and can be used as an in- memory data structure. The hash functions are selected in such a way that it avoids collision. Hashing as a searching technique is proven to be more apt for equality searches compared with tree structures that are better for range searches.

The hash functions create an index that helps to identify the documents that match the query posed by the user. A hash file called as a signature file is used for filtering. The filtering generally identifies the documents that match the query with a near precision. The filtering process uses the hash function and creates a signature for each of the document. An inverted file has a hash file that has a list of sorted words, each associated with a set pointers to the page in which it occurs. The authors of [1] list out the various data structures used for indexing in the information retrieval process. They discuss the effectiveness of hashing, B trees, B+ trees for implementing the index structures.

## 6. Process-oriented Data structures in Information retrieval

A stack is a linear data structure which uses one end of the data structure for storage and retrieval of data items. A stack is used in information retrieval algorithms for string matching in suffix arrays.

A graph is a data structure with nodes and edges connecting. It is one of the data structures that find wide application in multiple fields. It has been used to find the relationship between two data items or components or to find the connectivity between two different nodes in computer network.

In information retrieval, graphs are used to find the relationship between the queries posed by the user and the documents present in the corpus. The implementation of semantic nets and frames look into the similarities in the structure of query and document graphs. Graphs are also used to define the search space in the field of information retrieval. The graph structures frame the base for concept networks used in fuzzy information retrieval. Each node represents a concept or a document. In a concept network, an edge is used to connect two different concepts, $C_i$, to a document, $D_i$, and the edge is labeled with a real value between zero and one which signifies the fuzzy weightage given to the relationship.

The graphs are also used in web based information retrieval for providing relevance score based on the relevance propagation in document graph [9]. The other areas where the graphs find an application in the field of information retrieval are collaborative filtering, classification of the retrieved documents and unified link analysis.

## 7. Descriptive data structures in information retrieval

A tree is a data structure has a data item as its root and the subtrees are generated with a node as a parent node. In a search tree generally the solution is found as leaf. There are various kinds of trees depending upon the way of arrangement and the way of traversal of the tree. A B tree is a binary search tree that has an additional property of self – balancing. The advantage of B-tree is that the searching needs only a logarithmic time. A B+ tree is a self- balancing tree that can adjust its height and the nodes are linked used pointers. These pointers enable the range searches to be implemented efficiently in a B+ tree structure. A digital tree is one where the left subtree is traversed for a 0 and the right tree is traversed for a bit value of 1.

Generally any search operation can be treated as a generation of node in a search tree. The binary tree structures like B trees, B+ trees are used for implementing the index in the information retrieval process. A B-tree is used for implemented the inverted files. A Prefix B-tree is does not store the entire prefixes, but the tree is reconstructed each time the tree is searched [5]. This particular method provides the advantages of B-trees, digital search trees, and key compression techniques. It also reduces the processing overhead of compression techniques.

A trie is a data structure that is used for storing a string starting from the root node and proceeding till the leaf node. Figure 2 is an excerpt from [6] where the authors show the storage of strings, ape, apple, organ and organism in a trie.

A PAT tree in the field of information retrieval is a binary tree structure where PAT is a short form for PATRICIA, which in turn is an abbreviation for "Practical Algorithm to Retrieve Information Coded in Alphanumeric". A PAT is a simple variant on a trie in which any path whose interior vertices all have only one child is compressed into a single edge. It is a trie data structure with radix of 2, meaning that each bit of the key is compared individually and each node is a two-way (i.e., left versus right) branch. It differs from the tries in that Patricia trees contain no nodes with only one child. Every node is either a leaf or has at least two children. This immediately implies that the number of internal (non-leaf) nodes does not exceed the number of leaves.

Linear linked lists are data structures that help in efficient traversing of data items. It is a amalgamation of a set of data objects with pointers or references to the next data item. If it is a doubly linked list we would have pointers to the preceding data item and also to the succeeding data item. Linear linked lists are used for implementing posting lists.

A posting list is a data structure that maintains the list of documents that contains a particular term. Generally a dictionary of terms is built and then for each term in the dictionary a posting list is formed containing the list of documents that contains the particular term. To traverse a posting list again another data structure, pointer, is used which is called as a skip pointer. A pointer is generally a variable which holds the address of a data item.

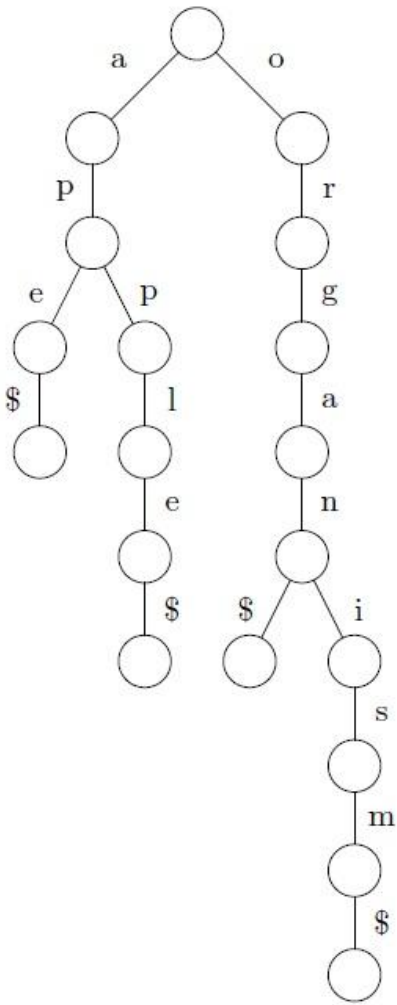

**Figure 2: A Trie storing a string**